\pdfoutput=1
\documentclass{PoS}
\title{Searches for very rare decays to purely leptonic final states at LHCb}

\ShortTitle{Searches for very rare decays to purely leptonic final states at LHCb}

\author{\speaker{Mathieu Perrin-Terrin}%
         \thanks{on behalf of the LHCb Collaboration.}\\
        CPPM, Aix-Marseille Universit\'e, CNRS/IN2P3, Marseille, France\\
        E-mail: \email{perrin-terrin@cppm.in2p3.fr}}

\abstract{We present a review of the searches for very rare decays to muonic final states performed at LHCb using 1.0 \invfb\ of \pp\ collisions at 7 \TeV\ centre of mass energy. Flavour changing neutral current processes, such as \bqtomumu\ and \bqtommmm\ are highly suppressed in the Standard Model (SM). Such decays therefore allow contributions from new processes or new heavy particles to significantly modify the expected SM rates. Charged lepton flavour violating processes, such as the neutrino-less \tautommm\ decay, have vanishingly small decay rates in the SM, but can be significantly enhanced in extended models. We report the latest results on these channels from LHCb. }

\FullConference{36th International Conference on High Energy Physics,\\
		July 4-11, 2012\\
		Melbourne, Australia}

\newcommand{\taummm}{\decay{\tau^-}{\mu^-\mu^-\mu^+}}

\newcommand{\DPhipi}{\decay{\Dsm}{\Phi(\mumu)\pi^-}}

\newcommand{\tautommm}{\taummm}

\newcommand{\Dsetammmnu}{\decay{\Dsm}{\eta(\mumu\gamma)\mu^-\BAR{\nu}_\mu}}
\newcommand{\bqtommmm}{\Bqmumumumu}

\newcommand{\bstommmma}{\Bsmumumumu}
\newcommand{\bdtommmma}{\Bdmumumumu}

\newcommand{\Bmumu}{\Bqmumu}

\newcommand{\bdtomumua}{\Bdmumu}
\newcommand{\bstomumua}{\Bsmumu}
\newcommand{\bqtomumu}{\Bqmumu}

\newcommand{\BsJpsiPhi}{\decay{\Bs}{\Jphi}}
\newcommand{\BsJpsiPhiMuons}{\decay{\Bs}{\Jpsi(\mumu)\Phi(\mumu)}}

\newcommand{\BKp}{\decay{\Bd}{\Kplus\pim}}

\newcommand{\BuJpsiK}{\decay{\Bu}{\Jpsi \Kplus}}

\newcommand{\CL}{C.L.\ }
\newcommand{\CLsb}{\ensuremath{\textrm{CL}_{\textrm{s+b}}}}
\newcommand{\CLs}{\ensuremath{\textrm{CL}_{\textrm{s}}}}
\newcommand{\CLb}{\ensuremath{\textrm{CL}_{\textrm{b}}}}
\newcommand{\comment}[1]{\par\noindent {\em\small [#1]}}
\renewcommand{\comment}[1]{}

\newcommand{\unit}[1]{\ensuremath{\rm\,#1}}

\newcommand{\GeV}{\unit{GeV}}

\newcommand{\TeV}{\unit{TeV}}
\newcommand{\invfb}{\unit{fb^{-1}}}

\newcommand{\BF}{\ensuremath{{\cal B}}}
\newcommand{\BRof}[1]{\ensuremath{{\cal B}(#1)}}

\newcommand{\BAR}[1]{\overline{#1}}

\newcommand{\particle}[1]{{\ensuremath{\rm #1}}}

\newcommand{\pp}{\particle{pp}}

\newcommand{\Bd}{\particle{B^0}}
\newcommand{\Bs}{\particle{B^0_s}}
\newcommand{\Bds}{\particle{B^0_{(s)}}}
\newcommand{\Bu}{\particle{B^+}}

\newcommand{\Dsm}{\particle{D_s^-}}

\newcommand{\Bq}{\particle{B^0_{(s)}}}

\newcommand{\Jpsi}{\particle{J\!/\!\psi}}
\newcommand{\Jmm}{\particle{\Jpsi(\mu\mu)}}

\newcommand{\Kstbar}{\particle{\BAR{K}^{*0}}}  

\newcommand{\Kplus}{\particle{K^+}}

\newcommand{\pip}{\particle{\pi^+}}
\newcommand{\pim}{\particle{\pi^-}}

\newcommand{\Jphi}{\particle{J\!/\!\psi \phi}}

\newcommand{\decay}[2]{\particle{#1\!\to #2}}

\newcommand{\mumu}{\particle{\mu^+\mu^-}}

\newcommand{\Bhh}{\decay{\Bds}{h^+h'^-}}                  %         B{d,s}->hh
\newcommand{\BdJmmKS}{\decay{\Bd}{\Jmm\Kstbar(\Kplus\pi^-)}}              %  411300 Bd->Jpsi(mumu)KS  
\newcommand{\Bsmumu}{\decay{\Bs}{\mu^+\mu^-}}            %         Bs->mumu
\newcommand{\Bqmumu}{\decay{\Bq}{\mu^+\mu^-}}            %         Bs->mumu
\newcommand{\Bdmumu}{\decay{\Bd}{\mu^+\mu^-}}            %         Bs->mumu

\newcommand{\Bsmumumumu}{\decay{\Bs}{\mu^+\mu^-\mu^+\mu^-}}            %         Bs->mumumumu
\newcommand{\Bqmumumumu}{\decay{\Bq}{\mu^+\mu^-\mu^+\mu^-}}            %         Bs->mumumumu
\newcommand{\Bdmumumumu}{\decay{\Bd}{\mu^+\mu^-\mu^+\mu^-}}            %         Bs->mumumumu

\newcommand{\beq}{\begin{equation}}
\newcommand{\eeq}{\end{equation}}

\begin{document}
\section{Introduction}

% The three searches for \bqtomumu, \bqtommmm\ and \tautommm\ are presented in this paper with their phenomenological motivations. 
% These searches were performed by analysing the 1.0 \invfb\ of \pp\ collisions recorded in 2011 with the LHCb detector \cite{lhcb} installed at the Large Hadron Collider. The LHCb detector is a single harm forward spectrometer that covers the region of pseudorapidity between 2 to 5.

The searches for \bqtomumu, \bqtommmm\ and \tautommm\ performed 
by analysing 1.0~\invfb\ of \pp\ collisions recorded in 2011 with the LHCb detector \cite{lhcb} are presented.
% The LHCb detector is a single harm forward spectrometer that covers the region of pseudorapidity between 2 to 5.
The three decay modes being purely muonic, the analyses have similarities. They all take advantage of the efficient muonic trigger lines \cite{trigger}, that for instance allow to trigger events with a muon of low transverse momentum (down to 0.5 \GeV\ for di-muons), and reach an efficiency of 90\% for di-muon signal. In addition, high rejections of combinatorial backgrounds are achieved using the excellent tracking performances (e.g. the resolution on the momentum is $\frac{\delta p}{p} \in [0.4,0.6] \textrm{ for } p\in[5;100]\GeV$). Finally physical backgrounds, where hadrons are misidentified as muons, are reduced with the particles identification detectors that yield an efficiency for identifying muon of 97\% for misidentification rate of pions (kaons) into muons below 3 (5)\%.
To avoid unconscious bias, the events for which the $\Bq$ or $\tau$ mass candidates lie within three mass resolutions of the expected mean are kept hidden until all crucial choices for the analysis were made.

\section{Searches for \bqtomumu}

%\subsection{Motivations}
Within the Standard Model (SM), \bqtomumu\
%exclusive dimuon decays of the $B^0_{(s)}$ mesons 
are rare processes as they
occur only via loop diagrams and are helicity suppressed. 
The amplitudes contributing to the branching fraction (\BF) 
can be expressed in terms of the scalar ($c_S$), pseudoscalar ($c_P$) and axial vector ($c_A$) Wilson coefficients in a  general approach~\cite{Bobeth2001}. 
Within the SM, $c_{S}$ and $c_{P}$ contributions are negligible while $c_A$ is 
calculated with a few percent accuracy~\cite{SMprediction} and leads~\cite{Buras2010} to a prediction of  
$\BRof \bstomumua_{\rm SM}  =  (3.2 \pm 0.2) \times 10^{-9}$ and $\BRof \bdtomumua_{\rm SM}  =  (1.0 \pm 0.1) \times 10^{-10}$.
Models beyond the SM could contribute to these Wilson coefficients and change significantly the \BF. For instance, within the Minimal Supersymmetric SM 
(MSSM) in the large $\tan \beta$ approximation~\cite{MSSM}, $c_{S,P}^{\rm MSSM} \propto \tan^3\beta/M_A^2$, 
where $M_A$ denotes the pseudoscalar Higgs mass and $\tan \beta$ the ratio of Higgs vacuum expectation values.
The LHCb experiment sets the most restrictive upper limits~\cite{bsmm}, 
\BRof\bstomumua $<1.4 \times 10^{-8}$ and 
\BRof\bdtomumua $<3.2 \times 10^{-9}$ at 95\% \CL 
 The present paper expounds the update of this analysis using the full 2011 data set.\footnote{Since then, the analysis has been updated leading to the first evidence for the \Bsmumu decay~\cite{bsmmEvidence}.}

%\subsection{Analysis description}
The strategy of the analysis is to derive the expected numbers of events of background and signal for a given \BF\ hypothesis and to compare these numbers to the observed ones with the \CLs\ method~\cite{CLs}. 
%. The compatibilities of the observed data with the background only and signal plus background hypotheses are derived with the \CLs method~\cite{CLs} .
The method provides \CLsb (\CLb), a measure of the compatibility of the observed distribution with the signal plus background (background only) hypothesis, and $\CLs = \CLsb / \CLb $ which is used to set upper limits on the \BF.

%The number of events corresponding to a \BF\ hypothesis
For a given \BF, the number of events 
 is obtained by scaling the yields observed in three controls channels, \BsJpsiPhi, \BKp\ and \BuJpsiK, by the ratio of the \BF\ hypothesis to the control channel \BF. The scaling factor corrects also for the different efficiencies between signal and control channels and for different intial states using $f_s/f_d$ measured at LHCb~\cite{fsfd}.

%Corrections are added in the scaling factor to account for the different efficiencies and the different initial states
%by normalising the signal to the three channels, \BsJpsiPhi, \BKp\ and \BuJpsiK\ which have respectively similar initial %state, topology and final state, for which both the observed yields and the branching fraction are known.
% The normalisation accounts for the different efficiencies and initial state using $f_s/f_d$ measured at LHCb~\cite{fsfd}.

To improve the sensitivity of the analysis, a selection which reduces the backgrounds is applied. Backgrounds stem from combinatorics where the two muons come from two different b-mesons and from peaking background from \Bhh\ (h standing for \Kplus\ or \pip) where hadrons are identified as muons. The first type of background is reduced by cutting on topological and kinematical variables and on a combination of them obtained with a Boosted Decision Tree (BDT). This selection has an efficiency similar on the signal, normalisation and control channels.
%This selection is very efficient for the signal and has similar efficiency on the signal, normalisation and control channels.
% No mass cut are applied since this variable is used to select a sample of pure combinatorics background from which the number of e
The peaking background is reduced using information from the particle identification detectors.

\begin{figure}%[htb]
  \begin{center}
    \includegraphics*[width=0.33\textwidth]{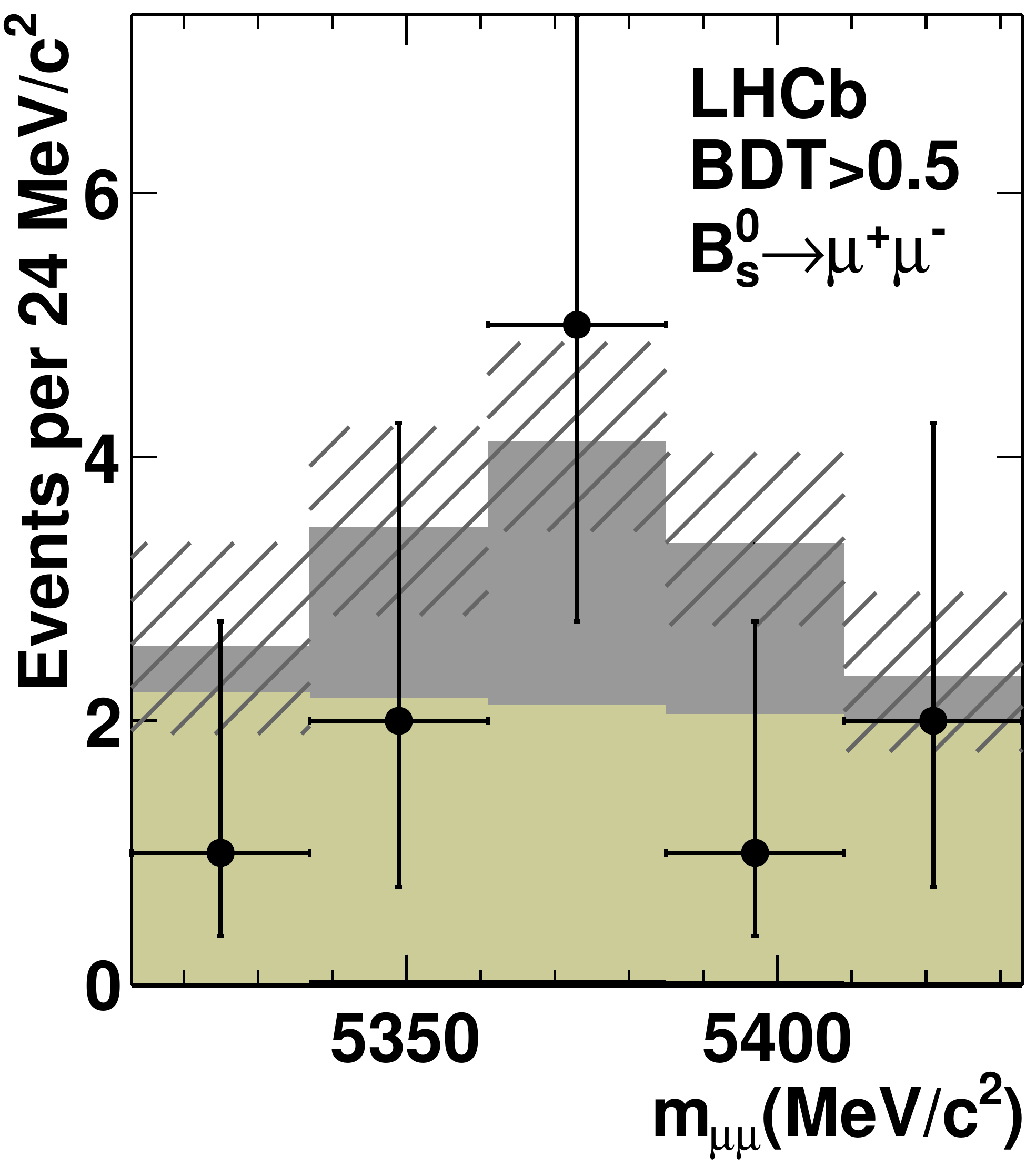}
    \includegraphics*[width=0.33\textwidth]{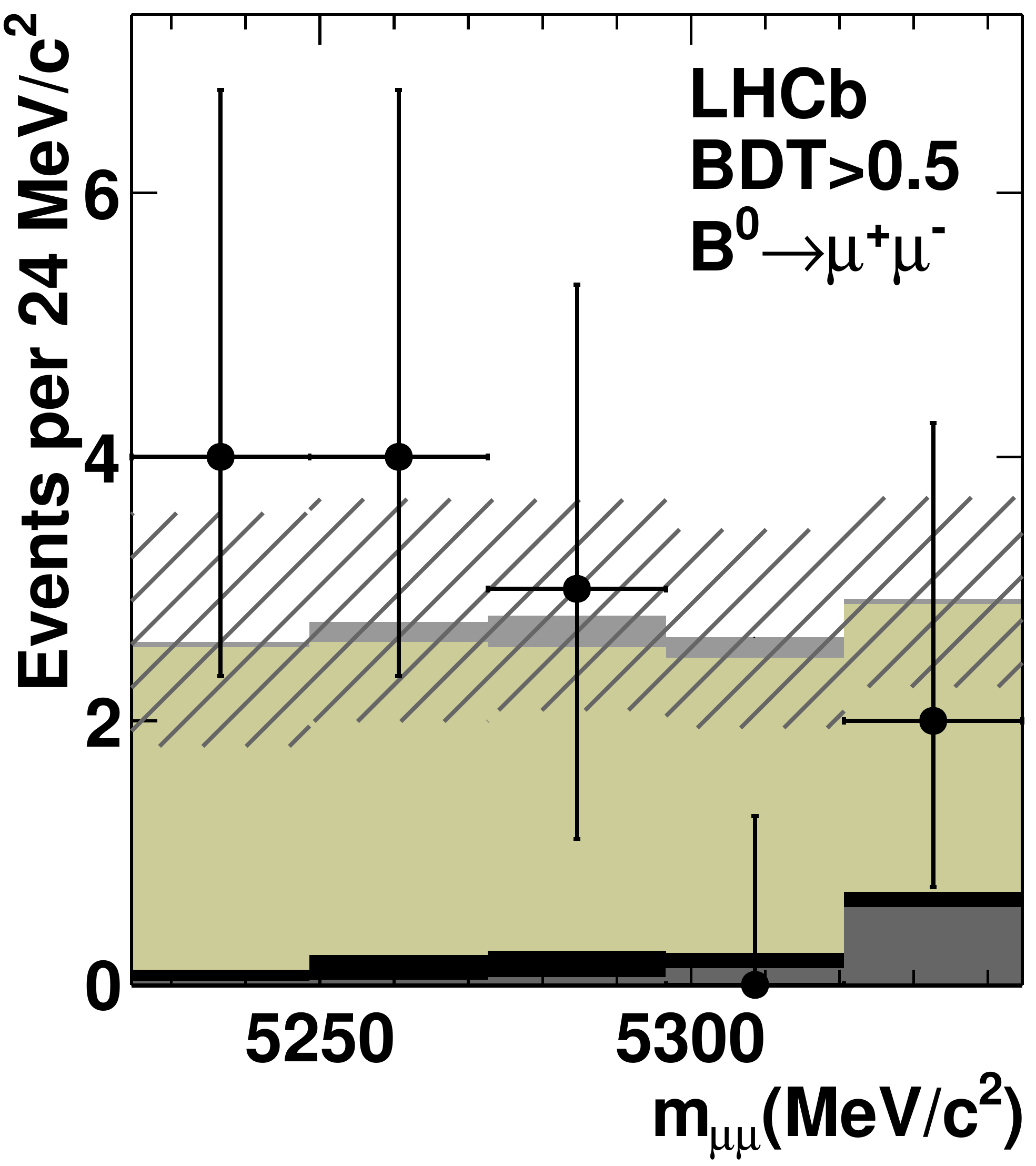}
  \end{center}
\vspace{-8mm}
\caption{Distribution of selected candidates (black points) in the (left) \Bsmumu\  and (right) \Bdmumu\ 
mass window for BDT$>$0.5, and expectations for, from the top, \Bsmumu\ SM signal (gray), combinatorial background (light gray), \Bhh\ background (black), and cross-feed of the two modes (dark gray). The hatched area depicts the uncertainty on the sum of the expected contributions.}
\label{fig:fondo_bsd}
\end{figure}

After this selection, events are classified in bins of the di-muon invariant mass and bins of a topological variable built with a second BDT. In each bin the expected numbers of signal and background events are derived.
% For the signal, the total number of events is obtained by normalising to the three mentioned channels.
 The signal mass shape is assumed to be a Crystal Ball shape, where the mean and the resolution are obtained on data while the transition point is derived from simulations. The BDT shape is obtained on data by extracting with a fit to the mass distribution, in each BDT bin, the yields of \Bhh.
For the combinatorial background, the mass and the BDT shape are obtained simultaneously by interpolating, in each BDT bin, the mass side-bands into the signal regions with an exponential function.
Finally, the peaking background BDT shape is assumed to be the same as the signal one 
(correcting for trigger bias) and the mass shape is taken from simulations. The total number of peaking backgrounds is derived by a data driven method.
The observed data and the expected distributions for the SM signal and background components are shown in Figure~\ref{fig:fondo_bsd}.

%\subsection{Results and Prospects}
% The compatibility of the observed distribution of events  
% with that expected for a given branching fraction 
% hypothesis is computed using the \CLs method~\cite{Read_02}.
% The method provides \CLsb, a measure of the 
% compatibility of the observed distribution with the signal plus background 
% hypothesis, \CLb, a measure of the compatibility with the background-only 
% hypothesis, and \mbox{$\CLs=\CLsb/\CLb$}.

\begin{figure*}%[!htb]
\centering
\includegraphics[width=0.45\textwidth]{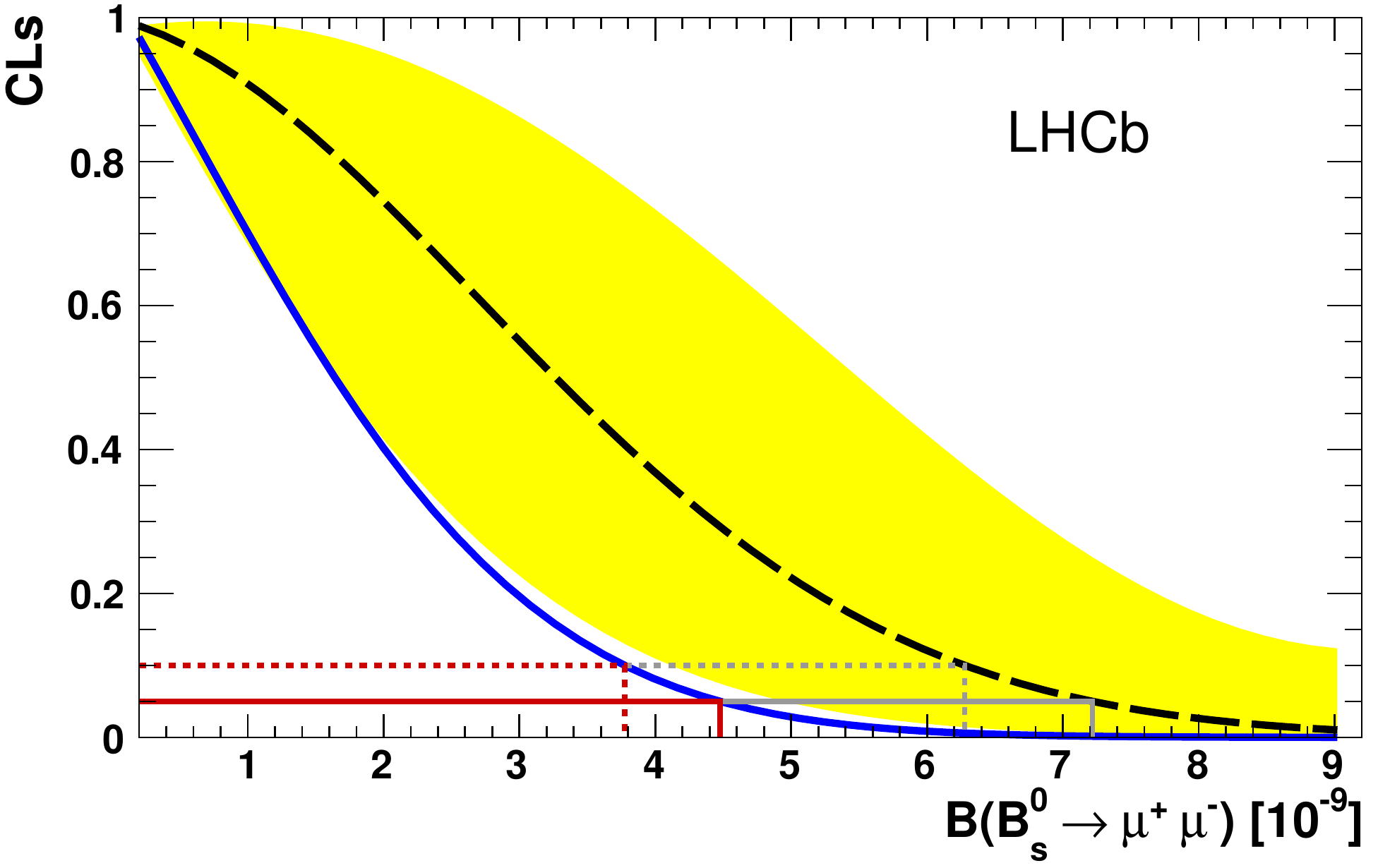}
\includegraphics[width=0.45\textwidth]{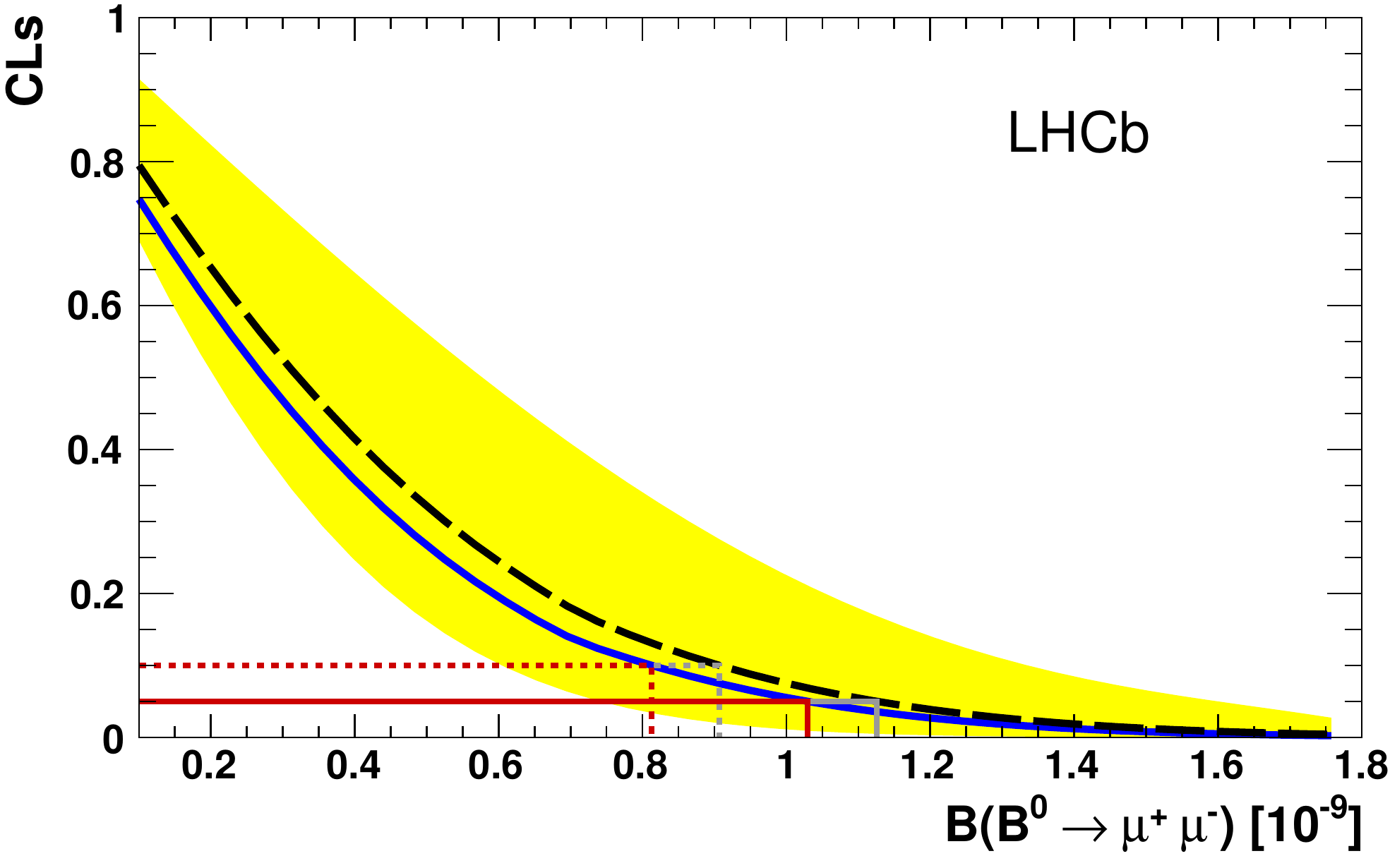}
\vspace{-4mm}
\caption
{
 \CLs\ as a function of the assumed \BF\ for (left) \Bsmumu and (right) 
\Bdmumu decays.
The long dashed black curves are the medians of the expected \CLs\ distributions for \Bsmumu, if background and SM signal were observed, and for \Bdmumu, if background only was observed.
The yellow  areas cover, for each \BF, 34\% of the expected \CLs\ distribution on each side of its median.
The solid blue curves are the observed \CLs. The upper limits at 90\,\% (95\,\%) \CL\ are indicated by the dotted (solid) horizontal lines in red (dark gray) for the observation and in gray for the expectation.}
\label{fig:cls_bsbd}
\end{figure*}

The comparison of the distributions of observed events and expected
background events results in a p-value \mbox{$(1-\CLb)$} of 18\,\% (60\,\%)
for the \Bsmumu\ \mbox{(\Bdmumu)} decay, where the \CLb\ values are those
corresponding to $\CLsb=0.5$.
No excess being observed, uppers limits are set and reported in Table~\ref{tab:bds_results}.
The expected and observed \CLs\ values are shown in Fig.~\ref{fig:cls_bsbd} for the \Bsmumu\ and \Bdmumu\ channels,
each as a function of the assumed \BF.
%The expected and measured limits for \Bsmumu and \Bdmumu at 90\,\% 
%and 95\,\% \CL\ are reported in Table~\ref{tab:bds_results}.

% The expected limits are computed allowing the 
% presence of \Bmumu events according to the SM branching fractions, including 
% cross-feed between the two modes. 

\begin{figure*}[!htb]
\centering
\includegraphics[width=0.45\textwidth]{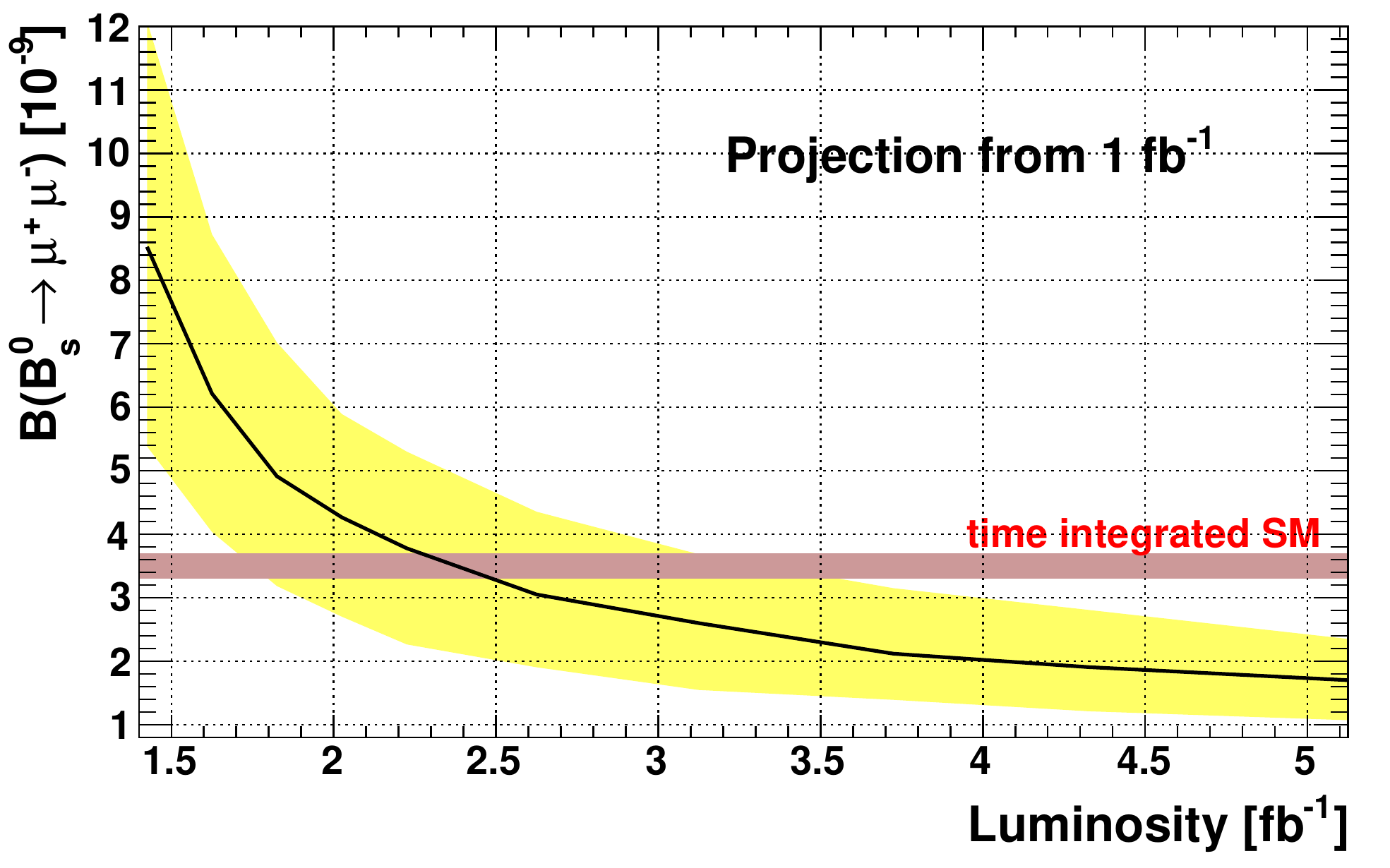}
\vspace{-4mm}
\caption
{\BF\ at which the \Bsmumu\ decay would be observed at 3$\sigma$ as a function of the integrated luminosity. The black plain line is the median of the \BF\ distribution giving a 3$\sigma$ observation. The yellow  areas cover, for each integrated luminosity, 34\% of the \BF\ distribution on each side of its median. The red horizontal band represents the time integrated SM prediction for \BRof\Bsmumu~\cite{BR_time_integrated}.}
\label{fig:prospect}
\end{figure*}

\begin{table}%[!htb]
\begin{center}
\small{
\begin{tabular}{@{}l|cc|cc@{}}
\hline\hline 
Mode &\multicolumn{2}{c}{\Bsmumu}&\multicolumn{2}{|c}{\Bdmumu}\\
          Limit & at 90\,\% \CL & at 95\,\% \CL& at 90\,\% \CL & at 95\,\% \CL\\
\hline 
       Exp. bkg+SM           &  $6.3 \times 10^{-9} $  & $ 7.2  \times 10^{-9} $&&\\ 
              Exp. bkg              &  $2.8 \times 10^{-9} $  & $ 3.4  \times 10^{-9} $ &  $0.91 \times 10^{-9}$  & $ 1.1 \times 10^{-9}$  \\ 
              Observed              &  $3.8 \times 10^{-9} $  & $ 4.5  \times 10^{-9} $ &                 $0.81 \times 10^{-9}$  & $ 1.0 \times 10^{-9}$  \\ 
% \hline
% \Bdmumu      & Exp. bkg              &  $0.91 \times 10^{-9}$  & $ 1.1 \times 10^{-9}$\\ 
%              & Observed              &  $0.81 \times 10^{-9}$  & $ 1.0 \times 10^{-9}$   \\ 
\hline \hline
\end{tabular}
\vspace{-1mm}
\caption{Expected and observed limits on \BRof\Bmumu. 
\label{tab:bds_results}
%The expected limits are computed allowing the presence of \Bmumu events 
%according to the SM branching fraction.
}}
\end{center}
\end{table}
Assuming the same sensitivity for the analysis, prospect for the discovery of the \Bsmumu\ decay can be derived. Figure~\ref{fig:prospect} shows the value of \BRof\Bsmumu\ for which a 3 $\sigma$ evidence of the decay could be obtained as a function of the integrated luminosity. This extrapolation takes into account the increase of energy at the center of mass to 8\TeV\ and the already observed data pattern for 2011.

\section{Searches for \bqtommmm}
The decays of \Bq\ mesons to four muons are strongly suppressed in the SM and, for \Bs, mainly occur via the resonant mode \BsJpsiPhiMuons\ for which the \BF\ is estimated to be $(2.3\pm0.9)\times 10^{-9}$~\cite{pdg}. In the SM, the \BF\ of the non-resonant modes are predicted not to exceed $10^{-10}$~\cite{b24mu_prediction}, but could receive non negligible contributions from models beyond the SM. % in which new particles would decay into a pair of muons. 
These decay modes have therefore potential for discovering physics beyond the SM and reassessing the hyper-CP anomaly~\cite{hyperCP}.

The searches are implemented with a cut and count analysis. Similarly to the \Bqmumu\ analysis, a control channel, \BdJmmKS, is used to derived the number of signal events corresponding to a \BF\ hypothesis. The selection is designed to be similar for the signal and the normalisation channel. The four muon candidates are required to be made of tracks with a good particle identification and to have an impact parameter significance (IPS) with respect to the primary vertex larger than 16. The candidate should also form a vertex with a $\chi^2$ per degree of freedom smaller than 6 and have an IPS smaller than 9. Resonant candidates are vetoed when their di-muons invariant masses lie around the masses of the $\Jpsi$ and $\Phi$.

\begin{figure*}%[!htb]
\centering
\includegraphics[width=0.45\textwidth]{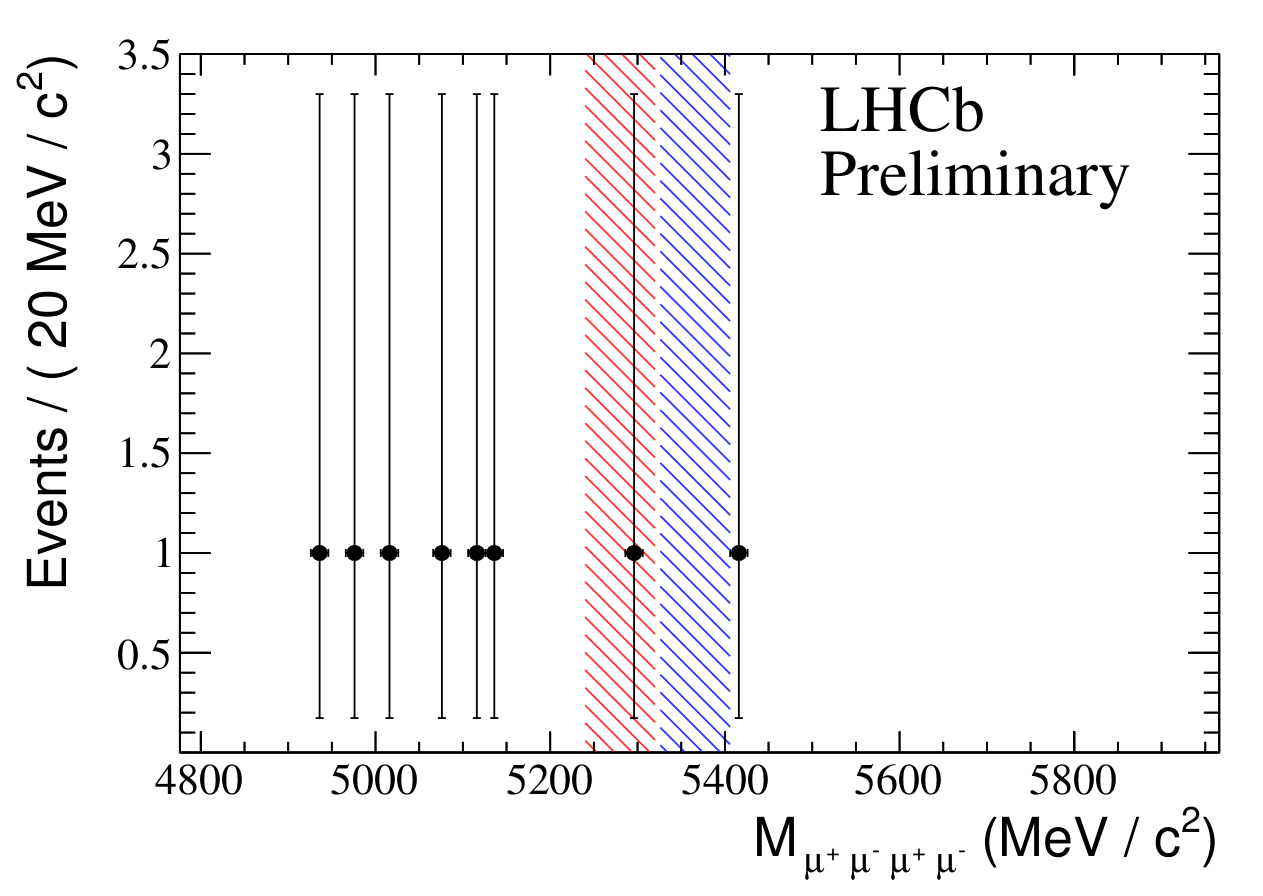}
\vspace{-4mm}
\caption
{Observed invariant mass distribution of the \bqtommmm\ candidates. The blue (red) dashed area is the \Bs (\Bd) mass window.}
\label{fig:b4m}
\end{figure*}

The only relevant source of background is the combinatorial one. The expected number of these events is obtained by interpolating the side-bands and yields to $0.38^{+0.23}_{-0.17}$ ($0.30^{+0.22}_{-0.20}$) events in the \Bs (\Bd) mass window. The observed pattern is shown in Figure~\ref{fig:b4m} and is compatible with the background expectation. Using the \CLs\ method~\cite{CLs} the first limits on these two processes are set to  $\BRof{\bstommmma} < 1.3\times 10^{-8}$ and $\BRof{\bdtommmma} < 5.4\times 10^{-9}$ at 95\% \CL.

% \begin{eqnarray*}
%    \BRof{\bstommmma} &<& 1.3\times 10^{-8},
%  \\\BRof{\bdtommmma} &<& 5.4\times 10^{-9}
% \end{eqnarray*}
% at 95\% \CL.
\section{Search for \tautommm}

In the SM the lepton flavour violation (LFV) decay \tautommm\ is almost completely suppressed, while many theories beyond the SM~\cite{t23mu_prediction} predict an enhanced LFV and in particular in the $\tau$ sector, raising in some cases \BRof\tautommm\ to an experimentally observable level. 

The analysis strategy is similar to the one used for the searches of \Bqmumu.
%  The main background processes consist of long decay chains of heavy mesons with three real muons in the final state or including one or two real muons in combination with one or two mis-identified particles. The dominant contribution in these processes is coming from \Dsetammmnu, and is treated separately for the others.
 The main background processes consist of long decay chains of heavy mesons with three real or misidentified muons in the final state. The dominant contribution in these processes is coming from \Dsetammmnu, and is treated separately.

After a loose selection based on the kinematics of the decay, the three muon candidates are classified in bins of invariant mass and of two multivariate operators combining information of the topology of the candidates and of the particle identification of the tracks used to form them. The three variables line shapes are determined for the signal with data driven methods and for the background by interpolating from the mass side-bands or with simulated data for \Dsetammmnu. The signal yield corresponding to a \BF\ hypothesis is obtained by normalising to \DPhipi.

\begin{figure*}%[!htb]
\centering
\includegraphics[height=0.3\textwidth]{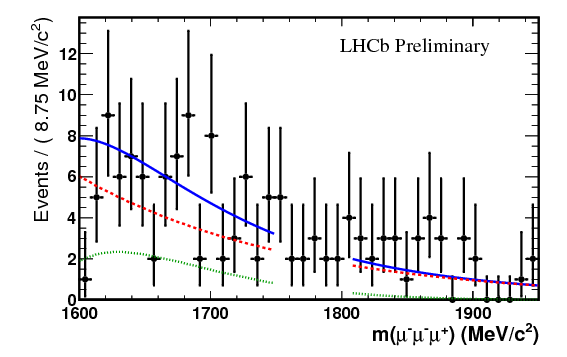}
\includegraphics[height=0.3\textwidth]{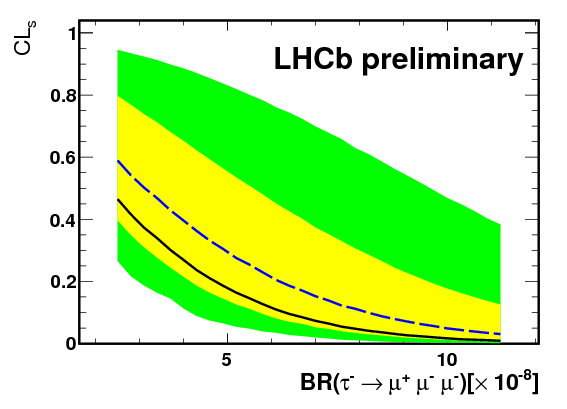}
\vspace{-4mm}
\caption
{
(left) Observed invariant mass distribution of the \taummm\ candidates in the most signal like merged bins. The solid line is the fit to the data of the sum of the combinatorial exponential contribution  (dashed red line) and of the \Dsetammmnu\ one (dashed green line).
(right) \CLs\ as a function of the assumed \BF. The long dashed blue curves are the medians of the expected \CLs\ distributions for \taummm\ if background only was observed.
The yellow (green) areas cover, for each \BF, 34\% (48\%) of the expected \CLs\ distribution on each side of its median.
The solid black curves are the observed \CLs.}
\label{fig:results_tmmm}
\end{figure*}

The observed data together with the background expectations are shown in Figure~\ref{fig:results_tmmm}. The compatibility of the data to the background or background plus signal hypothesis are computed with the \CLs\ method~\cite{CLs}. The observed data are compatible with the background only hypothesis. Figure~\ref{fig:results_tmmm} shows the \CLs\ as a function of the assumed \BF. An upper limit is set at $\BRof\taummm < 7.8 (6.3)\times 10^{-8}$ at 95 (90)\% \CL which is the first limit of a LFV $\tau$ decay set at a hadron collider.

\end{document}